\documentclass[english,11pt]{article}
\pdfoutput=1
\usepackage[a4paper]{geometry}
\usepackage{epsf,amsmath,amssymb,dcolumn}
\usepackage[pdftex]{graphicx}
\usepackage{scalefnt}
\usepackage{filemod}
\usepackage{subfig}
\usepackage{pstricks}
\usepackage{cite,hyperref}
\usepackage{color}
\usepackage{rotating}
\usepackage{xspace}
\usepackage{dsfont}
\usepackage{slashed}

\newcommand{\abbrev}{\scalefont{.9}}

\newcommand*{\ie}{i.e.\@\xspace}

\newcommand{\dd}[1]{\mathrm{d}#1\,}

\newcommand{\nn}{\nonumber\\}

\newcommand{\me}{\mathcal{M}}

\newcommand{\order}[1]{\mathcal{O}\left(#1\right)}

\newcommand{\lo}{{\abbrev LO}\xspace}
\newcommand{\nlo}{{\abbrev NLO}\xspace}
\newcommand{\nnlo}{{\abbrev NNLO}\xspace}
\newcommand{\ir}{{\abbrev IR}\xspace}
\newcommand{\uv}{{\abbrev UV}\xspace}
\newcommand{\dr}{{\abbrev DR}\xspace}
\newcommand{\qcd}{{\abbrev QCD}\xspace}

\newcommand{\fdr}{{\abbrev FDR}\xspace}
\newcommand{\barsqr}[1]{\overline{#1}^2}
\newcommand{\barfourth}[1]{\overline{#1}^4}

\newcommand{\deltafun}{\,\delta}
\newcommand{\thetafun}{\,\theta}
\newcommand{\dfun}{\,d}
\newcommand{\efun}{\,e}

\DeclareMathOperator{\adj}{adj}

\newcommand{\eqn}[1]{Eq.\,(\ref{#1})}
\newcommand{\neqn}[1]{Eqs.\,(\ref{#1})}
\newcommand{\fig}[1]{Fig.\,\ref{#1}}

\newcommand{\sct}[1]{Section~\ref{#1}}
\newcommand{\scts}[1]{Sections~\ref{#1}}

\newcommand{\refe}[1]{Ref.\,\cite{#1}}

\newcommand{\alphas}{\alpha_S}

\newcommand{\api}{\frac{\alpha_S}{\pi}}

\newcommand{\msbar}{\ensuremath{\overline{\text{MS}}}\xspace}

\newcommand{\mz}{M_Z}

\newcommand{\mw}{M_W}

\newcommand{\thw}{\theta_w}

\newcommand{\mt}{M_t}

\newcommand{\mtpole}{M_t^{\text{pole}}}

\newcommand{\mtmt}{M_t^{\msbar}}
\newcommand{\mtFDR}{M_t^{\text{FDR}}}
\newcommand{\mtbare}{M_{t}^{\text{bare}}}

\newcommand{\gamgam}{\gamma\gamma}

\newcommand{\GF}{G_F}

\newcommand{\veps}{\varepsilon}
\newcommand{\form}{\texttt{FORM}\xspace}
\newcommand{\mathem}{\texttt{Mathematica}\xspace}
\newcommand{\cpp}{\texttt{c++}\xspace}

\newcommand{\qgraf}{\texttt{qgraf}\xspace}
\newcommand{\qtoe}{\texttt{q2e}\xspace}
\newcommand{\expp}{\texttt{exp}\xspace}
\newcommand{\matad}{\texttt{MATAD}\xspace}

\newcommand{\cuba}{\texttt{CUBA}\xspace}

\newcommand{\jaxodraw}{\texttt{jaxodraw}\xspace}

\title{
  \begin{flushright}
    {\small 
      MPP-2015-301
    }
  \end{flushright}
\vspace*{2em} {\bf 
Numerical Evaluation of Two-Loop Integrals in FDR
}}
\author{ Tom J.E. Zirke\\[2em]
{\it Max Planck Institute for Physics}\\[0em]
{\it F\"ohringer Ring 6, 80805 M\"unchen, Germany}}
\date{}
\begin{document}
\maketitle

\vspace*{1cm}
\begin{abstract}
We present a method to numerically evaluate infrared-finite one- and two-loop integrals 
within the Four-Dimensional Regularization/Renormalization approach,
in which a small mass serves as regulator. 
Typical integrals exhibit a logarithmic dependence on this mass,
which we extract with the aid of suitable subtraction terms that can easily be integrated analytically
until the logarithmic structure is revealed.
As first physical applications to test the method, we calculate \qcd corrections to the decay rates of 
scalar and pseudoscalar Higgs bosons into two photons in the limit of an infinite top-quark mass 
as well as to the $\rho$~parameter.
\end{abstract}
\vfill

\section{Introduction}
Precise measurements at particle colliders such as the LHC necessitate theory predictions with 
competitive accuracy, which involve the evaluation of loop integrals.
Often one needs to go beyond the one-loop level,
for instance if the next-to-leading order (\nlo) accuracy is not sufficient 
or if the leading-order (\lo) diagrams are already loop-induced.
While the one-loop problem is solved in principle, the calculation of two-loop integrals is in general
still a major challenge with current technology. 
Thus any possible strategy to simplify loop calculations is highly appreciable. 

The Four-Dimensional Regularization/Renormalization (\fdr) approach proposed by Pittau 
in 2012~\cite{Pittau:2012zd}
could potentially provide a way to reduce the enormous effort required especially for next-to-next-to-leading 
order (\nnlo) calculations~\cite{Donati:2013voa}.
In this approach an additional small mass parameter $\mu$ is introduced, maintaining gauge invariance, however, 
and loop integrals are redefined in a way that they are finite in four dimensions.
Compared to working in Dimensional Regularization (\dr)~\cite{'tHooft:1972fi}, which requires the evaluation of 
$D=(4-2\epsilon)$-dimensional integrals, this offers several simplifications.
First of all the calculation of an $L$-loop observable does not require the knowledge of higher terms in the
$\epsilon$ expansion of the $(L-1)$-loop result, \ie no $\frac{\epsilon}{\epsilon}$ terms are generated.
In addition, the four-dimensionality avoids problems regarding the continuation of $\gamma_5$
to $D$ dimensions\footnote{Already in \refe{Pittau:2012zd} it was shown that \fdr produces the correct chiral 
anomaly at one-loop order.}
and, in principle, welcomes numerical approaches. 
Another advantage of \fdr is that infrared (\ir) divergencies\footnote{By \ir divergencies we mean both 
soft and collinear singularities.} are regulated at the same time, 
which was demonstrated to work at \nlo using a small gluon mass in the phase-space 
integration~\cite{Pittau:2013qla} but still needs to be explored for higher orders~\cite{Page:2015zca}.
For \ir-finite observables, to which we will restrict ourselves in this paper, 
complete one-loop \cite{Dedes:2012hf,Donati:2013iya,Donati:2013bca} and two-loop~\cite{Donati:2013voa} 
calculations have been performed finding agreement with \dr results.

Whereas the computations mentioned above were performed analytically, 
we will pursue a numerical approach in this paper.
For dimensionally-regulated integrals 
it is absolutely necessary to expand in the regulator $\epsilon$ 
before the numerical integration can be performed.\footnote{Trying to evaluate integrals for small values 
of $\epsilon$ is not only hopeless because of the polynomial 
dependence on $\frac{1}{\epsilon}$, but it would not allow the regularization of ultraviolet (\uv) and 
\ir divergencies at the same time, either.} 
This can be achieved (at one-loop order) using local subtraction terms in loop momentum
space~\cite{Nagy:2003qn} or (at in principle arbitrary loop order) in
Feynman parameter space using sector decomposition~\cite{Binoth:2000ps,Heinrich:2008si}, for example.
For \fdr integrals, however, the divergencies merely lead to a logarithmic dependence on the regulator $\mu$.
If the expected asymptotic behavior is known, one could in principle try to evaluate the integrals directly
for small values of $\mu$, subtract the corresponding logarithmic terms from the result, 
and extrapolate to $\mu=0$.
Nevertheless, this would lead to sharply peaked integrands that are difficult to evaluate numerically.
An expansion in the regulator on the integrand level is thus very helpful, 
although not strictly necessary.
A possible way to find suitable local subtraction terms on the level
of Feynman parameters is the main topic of this paper.

The structure of the rest of this paper is the following: In Section~\ref{sec:formalism} we present our idea 
how to construct subtraction terms on the level of Feynman parameters, 
essentially by partly linearizing the denominator. 
Some physical applications of this idea are shown in Section~\ref{sec:Applications} 
before we give our conclusions in Section~\ref{sec:conclusions}.

\section{Formalism}\label{sec:formalism}
We begin this section with a brief review of the definition of the \fdr integral~\cite{Pittau:2012zd} 
in \sct{sec:FDRint} before we elaborate on our approach to construct subtraction terms 
for the one and two-loop cases in \sct{sec:extlogs}.\footnote{For a presentation in full detail 
see~\refe{Zirke:phd}.}

\subsection[The FDR Integral]{The \fdr Integral}
\label{sec:FDRint}
Consider a dimensionally regulated, \ir-finite 
$L$-loop integral
\begin{align}
\mathcal{I}^\text{DR} = \mu_R^{2\epsilon L}\prod_{k=1}^L \int \dd{^D l_k} J(\{l_i, q_i, m_i\},\epsilon ),
\end{align}
where $q_i$ and $m_i$ denote external momenta and internal masses, respectively, 
and $D=4-2\epsilon$.\footnote{Since we restrict ourselves to the treatment of \uv divergencies in this paper, 
we identify the scale required to keep the dimensionality with the renormalization scale $\mu_R$ from the 
beginning.}
The \fdr interpretation of such an integral is based on a splitting of the integrand of the form
\begin{align}
J = \lim_{\mu\to 0} \left[ J_V + J_F \right],
\end{align}
where a new scale $\mu$ is introduced, which is required to be lower than all the other scales 
in the problem and will be identified with the renormalization scale later.
The splitting is defined in such a way that $J_V$ sums up the contributions 
from divergent vacuum configurations, which are assumed to be unphysical.
Consequently, $J_F$ must contain all the information on the physical process and,
since it is free of \uv divergencies, it can be calculated in four dimensions.

The divergent vacuum integrals contained in the integral over $J_V$ 
depend only on $\mu$, so they either yield contributions proportional to powers of $\mu^{2}$, 
which vanish in the asymptotic limit $\mu\to 0$, constant terms,
or logarithms of the form $\ln^k\left(\frac{\mu^2}{\mu_R^2}\right)$, $k\in\{1,2,\cdots,L\}$.
The latter can be eliminated by setting $\mu=\mu_R$.
Thus, if the \fdr integral is defined as
\begin{align}\label{eq:FDRdiff}
\mathcal{I}^\text{FDR} 
&\equiv \lim_{\mu\rightarrow 0} \prod_{k=1}^L \int \dd{^4 l_k}
J_F (\{l_i, q_i, m_i\}, \mu)\bigg|_{\mu=\mu_R} \nn
&= \mathcal{I}^\text{DR} - \lim_{\mu\rightarrow 0} \prod_{k=1}^L \int \dd{^D l_k} 
J_V (\{l_i\},\mu,\epsilon)\bigg|_{\mu=\mu_R},
\end{align}
the difference between $\mathcal{I}^\text{FDR}$ and $\mathcal{I}^\text{DR}$ is a ($\epsilon$-dependent) constant.

To construct the splitting into $J_F$ and $J_V$ the prescription is to replace the loop momentum in 
every propagator by $l_i^2\to l_i^2-\mu^2 \equiv \barsqr{l_i}$\footnote{In this context, $l_i$ can also mean
a linear combination of loop momenta.}$^{,}$\footnote{For fermionic propagators, it was originally required 
in~\refe{Pittau:2012zd} to replace $\slashed{l}\rightarrow \slashed{l} - \mu$, but it is also valid to 
calculate the fermion traces first and replace $l^2 \rightarrow l^2 - \mu^2$ afterwards~\cite{Donati:2013iya}.
Throughout this paper we pursue the latter approach. }
and to apply the \emph{partial-fraction relations}
\begin{subequations}
  \label{eq:FDRpartfrac}
\begin{align}
  \frac{1}{\barsqr{l_i}-m_j^2} &= \frac{1}{\barsqr l_i} 
  \left( 1 + \frac {m_j^2} { \barsqr l_i - m_j^2} \right), \\
  \frac{1}{\barsqr{(l_i+q_j)}-m_j^2} &= \frac{1}{\barsqr l_i} 
  \left( 1 + \frac {m_j^2 - q_j^2 - 2 l_i\cdot q_j} { \barsqr {(l_i + q_j)} - m_j^2} \right)
  \label{eq:FDRpartfracmom}
\end{align}
\end{subequations}
repeatedly to the integrand, until all the terms are either divergent vacuum integrals or \uv-finite.
The former can be identified as part of $J_V$ and be dropped, 
the latter as part of $J_F$ and be evaluated in four dimensions.\footnote{The language used here and 
around \eqn{eq:FDRdiff} holds strictly only for the one-loop case.
At higher loop orders, divergent sub-integrals occur, which should be treated as a lower-order integral
with the remaining loop momenta interpreted as external momenta.
As a result, one has to drop vacuum integrals that are multiplied by a factorized non-vacuum integral.
This is concretized in \refe{Page:2015zca} as \emph{sub-} and \emph{global vacua}.}
In this way, the \uv divergencies contained in $\mathcal{I}^\text{DR}$ are traded in for \ir divergencies
related to vanishing $\mu$.

For gauge theories it is essential that the rules $l_i^2\to \barsqr{l_i}$ are applied to the numerator 
as well so that the additional mass scale $\mu$ does not spoil gauge invariance. 
In addition, the $\mu^2$ piece has to be treated exactly as the corresponding $l_i^2$ term
until all \uv~divergencies are subtracted, \ie one should distinguish different $\mu_i^2$ and
count them as $l_i^2$ when deciding by power counting how often \eqn{eq:FDRpartfrac} 
has to be applied. In this way, gauge symmetry is preserved.
The purpose of this is to ensure that \fdr and \dr results are related in a 
consistent and universal way. 
They are connected by finite renormalization constants~\cite{Donati:2013voa,Page:2015zca},
like results obtained in different renormalization schemes.

Tensor reduction is allowed in \fdr as well and can be performed in four dimensions, 
which generates fewer terms. However, the resulting factors of $l_i^2$ are not to be interpreted as
$l_i^2-\mu_i^2$ and can thus be canceled against suitable propagators only 
at the price of introducing $\mu_i^2$ terms in the numerator. 
Such terms can give finite contributions and must be treated carefully.\footnote{For details see 
\refe{Page:2015zca}, for example, where they are described as \emph{extra integrals}.}

\subsection{Construction of the Subtraction Terms}
\label{sec:extlogs}

\subsubsection{General Structure in Feynman Parameter Representation}
After isolating and subtracting the \uv divergencies, an $L$-loop \fdr integral typically has the form
\begin{align}
  \mathcal{I}&=  \prod_{j=1}^L \int \dd{^4 l_j}
  N\left(\{l_i,q_i,m_i\},\mu^2\right) 
  \cdot \prod_{k=1}^{N_x}D^{\alpha_k}(p_k,q_k,m_k)
  \cdot \prod_{n=1}^{N_y} D_0^{\beta_n}(p_n),
  \quad \alpha_i,\beta_i\in\mathds{N},
\end{align}
where $p_i$ and $q_i$ are linear combinations of the loop and external momenta, respectively, and 
we distinguish two types of propagators:
\begin{subequations}
\label{eq:propagators}
\begin{align}
  D^{-1}(p,q,m) &= (p+q)^2-m^2-\mu^2 +i\varepsilon, \\
  D^{-1}_0(p) &= p^2-\mu^2 +i\varepsilon.
\end{align}
\end{subequations}
We restrict ourselves to cases where one can find a momentum routing such that $q_i^2\neq m_i^2$
for all $i$, \ie we assume sufficiently good \ir behavior. The \ir regime is then governed 
entirely by the $D_0$-type propagators, which are generated in heaps from applying the partial 
fraction relations \eqref{eq:FDRpartfrac} but may also originate from massless lines of the graph 
under consideration.
Furthermore, $N$ denotes a non-trivial numerator including scalar products of loop and 
external momenta as well as powers of $\mu^2$.

Since we exclude \ir divergencies, the integral will have an asymptotic small-$\mu^2$ 
behavior of the form
\begin{align}
\mathcal{I} = \sum_{i=0}^L A_i \ln^i\left(\mu^2\right) + \order{\mu^2}.
\end{align}
The $\order{\mu^2}$ terms should be dropped according to the definition of the \fdr integral
\eqref{eq:FDRdiff}
so we are interested only in the logarithmic behavior for $\mu^2\to 0$, \ie in the coefficients
$A_0,\cdots,A_L$.

In order to find the origin of the divergencies that occur in this limit,
it turns out to be useful to introduce Feynman parameters first.
As we will see, in Feynman parameter space one can construct simpler integrals that 
possess the same small-$\mu^2$ behavior as the original integral and can serve as local subtraction 
terms. These auxiliary integrands can be integrated analytically over at least a subset of the parameters so that 
their logarithmic dependence on $\mu^2$ becomes explicit. 
The remaining integrations can then be performed numerically.

Denoting the Feynman parameters for the two classes of propagators defined in \eqn{eq:propagators} 
$x_i$ and $y_i$, respectively, one obtains integrals of the form
\begin{align}
  \label{eq:FDRparageneric}
  \mathcal{I} &= \prod_{i=1}^{N_x} \int_0^1 \dd{x_i}
  \prod_{j=1}^{N_y} \int_0^1 \dd{y_j}
  \deltafun\left(1-\sum_{k=1}^{N_x} x_k - \sum_{l=1}^{N_y} y_l\right)\nn
  &\quad\cdot  
  \frac {p(x_1,\cdots,x_{N_x},y_1,\cdots,y_{N_y})}
  {\left( b^T \adj(A) b + \det(A) c \right)^{N_1} \det(A)^{N_2}}.
\end{align}
The $L\times L$ matrix $A$, the $L$-dimensional vector $b$, and the scalar $c$
are obtained by expressing the sum of all propagators in terms of the 
loop momenta $l=(l_1,\cdots,l_L)$:
\begin{align}
  \sum_{j=1}^{N_x}D(p_j,q_j,m_j) + \sum_{k=1}^{N_y} D_0(p_k)
  &= l^T A l + 2 b \cdot l -c.
\end{align}
The elements of $A$ are sums of $x_i$ and $y_i$ parameters,
$b$ contains terms of the form $x_i q_i$, and $c$ reads
\begin{align}
  \label{eq:FDRDelta}
  c &= \sum_{i=1}^{N_x} x_i (m_i^2-q_i^2) + \mu^2 - i\varepsilon.
\end{align}
$p$ is a polynomial whose coefficients depend on the kinematic invariants and is a result of 
higher powers of propagators as well as possible non-trivial denominators.
One way to treat the latter is to interpret them as inverse propagators and 
to calculate the derivative with respect to the corresponding Feynman parameter rather than integrating over 
it~\cite{Smirnov:2006ry}.

If we ask which region in parameter space the logarithmic divergencies regulated by $\mu$ 
originate from, it must be the region where the denominator of \eqn{eq:FDRparageneric}
is of order $\mu^2$.
Note that the only place where $\mu^2$ enters the denominator is $c$.
Assuming $q_i^2<m_i^2$ for all $i$,\footnote{If this condition is relaxed, threshold singularities will occur, 
which are usually avoided in the numerical integration by introducing a suitable deformation of the 
integration contour~\cite{Soper:1999xk,Nagy:2006xy}. 
In principle this appears to be possible for the method presented here, 
but it is beyond the scope of this paper.}
it follows that all the $x_i$ need to be small
in order to make the denominator of order $\mu^2$.
The term $b^T \adj(A) b$ vanishes as well in this limit, so that indeed 
$b^T \adj(A) b + \det(A) c \propto \mu^2$.

A potential problem arises if $\det(A)$ vanishes as well.
In a more detailed analysis, one finds that this produces overlapping singularities that need to 
be disentangled.
After discussing in brief the one-loop case, where this problem is absent,
we will show how this can be resolved for the case $L=2$.

\subsubsection{Subtraction Terms for the One-Loop Case} 
For $L=1$ a major simplification occurs: It is $\det(A)=\sum_{i=1}^{N_x} x_i + \sum_{i=1}^{N_y} y_i=1$,
which obviously never vanishes. In addition,
there is only one $y$ parameter, which we integrate out using the delta function to obtain
\begin{align}
 \mathcal{I}^{(1l)} &=  \prod_{i=1}^{N_x} \int_0^1 \dd{x_i} \thetafun\left(1-\sum_{k=1}^{N_x} x_k\right)
 \frac{ p\left(x_1,\cdots,x_{N_x},1-\sum_{k=1}^{N_x} x_k \right)}
  {\left( \mu^2 + \sum_{j=1}^{N_x} x_j (m_j^2-q_j^2) + \left(\sum_{j=1}^{N_x}x_j q_j\right)^2 \right)^{N_1}}.
\end{align}

As pointed out above, the region of interest is where all $x_i$ go to zero.
Applying the useful relation
\begin{align}
  \label{eq:scalrel}
  &\prod_{i=1}^n \int_0^1 \dd{x_i} \thetafun\left(1-\sum_{k=1}^n x_k\right)  f(x_1,\cdots,x_n) \nn
  = &\prod_{i=1}^n \int_0^1 \dd{x_i} \deltafun\left(1 -\sum_{k=1}^n x_k\right)
  \int_0^1 \dd{r} r^{n-1} f(r x_1,\cdots,r x_n),
\end{align}
one can achieve that this region is associated with the vanishing of only one parameter, namely $r$:
\begin{align}
  \label{eq:Intr1l}
 \mathcal{I}^{(1l)} &=  \prod_{i=1}^{N_x} \int_0^1 \dd{x_i} \deltafun\left(1-\sum_{k=1}^{N_x} x_k\right)
 \int_0^1 \dd{r} r^{N_x-1} \nonumber \\
 &\quad\cdot \frac{ p\left(r x_1,\cdots,r x_{N_x},1-r \right)}
  {\left( \mu^2 + r \sum_{j=1}^{N_x} x_j (m_j^2-q_j^2) + r^2  \left(\sum_{j=1}^{N_x}x_j q_j\right)^2 
    \right)^{N_1}}
\end{align}

Now we are ready to construct an auxiliary integrand that is easier to integrate but has the
correct dependence on $\ln(\mu^2)$.
For this purpose, we rewrite \eqn{eq:Intr1l} as
\begin{align}
 \mathcal{I}^{(1l)} &= \frac{1}{(M^2)^{N_1}}  
 \prod_{i=1}^{N_x} \int_0^1 \dd{x_i}
\deltafun\left(1-\sum_{k=1}^{N_x} x_k\right) \int_0^1 \dd{r} \tilde p(x_1,\cdots,x_{N_x},r) \nn
&\quad\cdot I_1^{(N_1,n_1,n_2,n_3)} \left(\frac{\mu^2}{M^2}, \frac{ \sum_{j=1}^{N_x} x_j (m_j^2-q_j^2)}{M^2},  
\frac{\left(\sum_{j=1}^{N_x}x_j q_j\right)^2}{M^2}; r \right),
\end{align}
where we have introduced a generic notation for the integrand,
\begin{align}
\label{eq:int1}
I_1^{(N_1,n_1,n_2,n_3)}(a, c_1, c_2; r)\equiv\frac{r^{N_1-1+n_1} (1-r)^{n_2} a^{n_3}}{(a + c_1 r + c_2 r^2)^{N_1}},
\end{align}
and factorized as many factors of $r$,$(1-r)$, and $a$ from $p$ so that the remainder $\tilde p$ is 
still a polynomial.

The behavior for small $\mu^2$ is now completely determined by the parameters of $I_1$.
In the case $n_1=n_3=0$, for example, there will be a logarithmic dependence on $\mu^2$.
We observe that the coefficient of $\ln(\mu^2)$ does not depend on $c_2$, \ie stetting $c_2=0$
would alter only the finite term.
Similarly, $n_2$ and the $r$ dependence of $\tilde p$ do not influence the logarithm.
Thus it suffices to calculate the auxiliary integral
\begin{align}
 \mathcal{A}^{(1l)} &\equiv \frac{1}{(M^2)^{N_1}}  
 \prod_{i=1}^{N_x} \int_0^1 \dd{x_i}
\deltafun\left(1-\sum_{k=1}^{N_x} x_k\right) \nn
&\quad\cdot \int_0^1 \dd{r} 
I_1^{(N_1,0,0,0)} \left(\frac{\mu^2}{M^2}, \frac{ \sum_{j=1}^{N_x} x_j (m_j^2-q_j^2)}{M^2},0; r \right)
\tilde p(x_1,\cdots,x_{N_x},0),
\end{align}
in order to reproduce the correct $\mu^2$ dependence in this case.
The denominator is now linear in $r$, which is the decisive simplification.
To get the correct finite part one now only has to calculate
\begin{align}
\lim_{\mu\to 0}\mathcal{R}^{(1l)}
&\equiv \lim_{\mu\to 0}\left\{\mathcal{I}^{(1l)} - \mathcal{A}^{(1l)} \right\}\nn
&= \frac{1}{(M^2)^{N_1}}  
 \prod_{i=1}^{N_x} \int_0^1 \dd{x_i}
\deltafun\left(1-\sum_{k=1}^{N_x} x_k\right) \int_0^1 \dd{r} \nn
&\quad\cdot \left\{
I_1^{(N_1,n_1,n_2,n_3)} \left(0, \frac{ \sum_{j=1}^{N_x} x_j (m_j^2-q_j^2)}{M^2},  
\frac{\left(\sum_{j=1}^{N_x}x_j q_j\right)^2}{M^2}; r \right)
\tilde p(x_1,\cdots,x_{N_x},r)\right. \nn
&\qquad \left. - 
I_1^{(N_1,0,0,0)} \left(0, \frac{ \sum_{j=1}^{N_x} x_j (m_j^2-q_j^2)}{M^2},0; r \right)
\tilde p(x_1,\cdots,x_{N_x},0) \right\},
\end{align}
where we were able to interchange the limit $\mu\to 0$ with the integration because
the integrand is well-behaved in the region of small $r$ by construction.
Thus the evaluation of the difference can be done completely numerically if necessary.

\subsubsection{Subtraction Terms for the Two-Loop Case} 
In the case $L=2$, one can label the momenta in such a way that 
each propagator contains either $l_1$, $l_2$, or $l_{12}\equiv l_1+l_2$.
We will distinguish three subclasses of propagators, depending on which of the three
momenta they contain. 
For the $D$-type propagators (cf. \eqn{eq:propagators}) we introduce three disjoint index sets 
$\mathcal{X}_1$, $\mathcal{X}_2$, and $\mathcal{X}_{12}$
with the following property: If a propagator contains $l_i$ ($i\in\{1,2,12\}$), 
the index of its Feynman parameter will be an element of $\mathcal{X}_i$.
Since the $D_0$-type propagators are uniquely determined by the loop momentum, 
we can simply define the parameter of $\frac{1}{l_i^2-\mu^2}$ to be $y_i$ for $i\in\{1,2,12\}$.

Using these conventions and assuming that the integral under consideration contains
at least one $D$ and one $D_0$ propagator with each $l_i$,\footnote{This provides the most complicated case. 
The other cases tend to be simpler but need to be distinguished carefully, which is done in detail in 
\refe{Zirke:phd}.} one can verify easily that
\begin{subequations}
\begin{align}
  \det(A) &= a_1a_2 + a_1a_{12} + a_2a_{12}, \\
  b^T \adj(A) b &= a_{12} (b_1 - b_2)^2 + a_2 (b_1 + b_{12})^2 + a_1 (b_2 +  b_{12})^2,
\end{align}
\end{subequations}
where
\begin{align}
\label{eq:defaibi}
a_i = y_i + \sum_{k\in\mathcal{X}_i} x_k, \quad 
b_i = \sum_{k\in\mathcal{X}_i} x_k q_k, \quad i\in\{1,2,12\}.
\end{align}

Now it is crucial to understand when a zero of $\det(A)$ overlaps with 
the region we are interested in, \ie where all $x_i$ are small.
Since it is
\begin{align}
  a_1 + a_2 + a_{12} = y_1+y_2+y_{12} + \sum_{i=1}^{N_x} x_i = 1
\end{align}
due to the delta function,
the point $a_1=a_2=a_{12}=0$ is outside the integration boundaries.
Thus only the zeros at $(a_1,a_2,a_{12})=(1,0,0),(0,1,0),(0,0,1)$ remain,
which, in view of \eqn{eq:defaibi}, are associated with zeros of either $y_1$, $y_2$, or $y_{12}$.
It turns out to be sufficient to factorize the worst-behaved zero.
Consider a typical structure that will lead to a logarithmic dependence on $\mu^2$:
\begin{align}
  \label{eq:FDRln2structure}
  \frac{1}{\barfourth{l_1}\barsqr{l_2}\barsqr{l_{12}}}
\end{align}
Since the propagator $\frac{1}{\barsqr{l_1}}$ is squared, its parameter $y_1$ will appear as an
extra factor in the numerator.
Thus the behavior at $a_1=1$ is worst because in the other cases this extra factor becomes small and 
partly compensates the vanishing of the denominator.

The next step is to find a parametrization that not only factorizes the worst zero of $\det(A)$
but also makes it possible to judge which terms can be neglected in an easy-to-integrate auxiliary 
integrand.
It is convenient to eliminate the parameter of the $D_0$-type propagator with the highest power,
which, like in the example above, we assume to be $y_1$ in the following.
To parametrize the zero of $\det(A)$ and the simultaneous vanishing of the $x_i$ in terms of a smaller set
of parameters,
we make use of the following transformation, which corresponds to applying \eqn{eq:scalrel} 
three times to different subsets of parameters:
\begin{subequations}
\label{eq:FDRfinaltrafo}
\begin{align}
x_i &\rightarrow (1-r) t x_i &\quad i\in\mathcal{X}_1, \\
x_i &\rightarrow rs x_i &\quad i\in\mathcal{X}_2\cup\mathcal{X}_{12}, \\
y_1 &\rightarrow (1-r)(1-t), &\\
y_i &\rightarrow r(1-s)  y_i &\quad i\in\{2,12\}.
\end{align}
\end{subequations}
The limit $(a_1,a_2,a_{12})\to(1,0,0)$ is mapped to $r\to 0$, and 
a factor of $r$ can be split off from $\det(A)$ and also from the complete denominator.
 
Application of this transformation to \eqn{eq:FDRparageneric} (reduced to $L=2$ and $N_y=3$) 
yields
\begin{align}
\label{eq:FDR2lfinal1}
  \mathcal{I}^{(2l)}
&= \prod_{i=1}^{N_x} \int_0^1 \dd{x_i}\deltafun\left(1-\sum_{l\in\mathcal{X}_1} x_l \right)
\deltafun\left(1-\sum_{m\in\mathcal{X}_2\cup\mathcal{X}_{12}} x_m \right) \nn
&\quad\cdot \int_0^1 \dd{y_2} \int_0^1 \dd{y_{12}} \deltafun\left(1-y_2-y_{12} \right)
 \int_0^1 \dd{r} \int_0^1 \dd{s} \int_0^1 \dd{t} I_2^{(N_1,N_2)}(r,s,t),
\end{align}
where again the behavior for small $\mu^2$ is determined by an integral over fewer parameters, namely
\begin{align}
\label{eq:FDR2lfinal2}
&I_2^{(N_1,N_2)}(r,s,t) \nn
&\equiv \frac{\tilde p (x_1,\cdots,x_{N_x},y_2,y_{12},r,s,t)}
 {\left[\left(1-r + r \dfun(s)\right) \left(\mu^2 + c_1 rs +c_2 (1-r) t\right) 
   + e(r,s,t) \right]^{N_1} \left(1-r + r \dfun(s)\right)^{N_2}}.
\end{align}
The coefficients
\begin{subequations}
\label{eq:FDRc12}
\begin{align}
c_1 &=  \sum_{k\in\mathcal{X}_2\cup\mathcal{X}_{12}} x_k (m_k^2 - q_k^2), \\
c_2 &=  \sum_{k\in\mathcal{X}_1} x_k (m_k^2 - q_k^2)
\end{align}
\end{subequations}
are constants with respect to $r$, $s$, and $t$, whereas $d$ and $e$ are polynomials in 
these variables:
\begin{subequations}
\begin{align}
\dfun(s) &= a_2(s) a_{12}(s), \label{eq:FDRintdfun}\\
\efun(r,s,t) &= (1-r)^2 t^2 b_1^2 + 2 rs(1-r)t \left[a_2(s) b_{12} - a_{12}(s) b_2\right]\cdot b_1 \nn
&\quad + r^2 s^2 \left[  a_2(s) b_{12}^2 + a_{12}(s) b_2^2 \right] + r s^2 (1-r) (b_2 + b_{12})^2,
\label{eq:FDRintefun}
\end{align}
\end{subequations}
where $b_i$, $i\in\{1,2,12\}$ as in \eqn{eq:defaibi} and 
\begin{subequations}
\label{FDRintas23}
\begin{align}
a_2(s) &=  (1-s) y_2 + s \sum_{k\in\mathcal{X}_2} x_k, \\
a_{12}(s) &=  (1-s) y_{12} + s \sum_{k\in\mathcal{X}_{12}} x_k.
\end{align}
\end{subequations}

Taking a closer look at \eqn{eq:FDR2lfinal2}, we see that the denominator is dominated by $\mu^2$
if $rs$ and $(1-r)t$ both vanish, \ie the logarithmic dependence is associated 
with three parameters.\footnote{In special cases, for which we refer the reader again to \refe{Zirke:phd}, 
not all of the three parameters $r$, $s$, and $t$ need to be present.}
In the limit $r\to 1$, which in view of \neqn{eq:defaibi} and \eqref{eq:FDRfinaltrafo} corresponds
to $a_1\to 0$, the integrand should be integrable because we required the worst, \ie logarithmic
divergency to be at $a_1 \to 1$. The balance between numerator and denominator both vanishing 
in this limit will still be present after the transformation of variables.

To eventually construct the auxiliary integrand, we replace the denominator
\begin{align}
D= \left(1-r + r \dfun(s)\right) \left(\mu^2 + c_1 rs +c_2 (1-r) t\right) + e(r,s,t)
\end{align}
by 
\begin{align}
\label{eq:Denoapprox}
\tilde D &= 
\left(1-r + r \dfun(0)\right) \left(\mu^2 + c_1 rs +c_2 (1-r) t + r s^2 (b_2 + b_3)^2 \right)\\
 &= D + \order{rs\mu^2,r^2s^2,rs(1-r)t,(1-r)^2t^2}.
\end{align}
In the numerator one should drop terms of $\order{rs\mu^2,r^2s^2,rs(1-r)t,(1-r)^2t^2}$ as well,
but this must be done in such a way that the behavior in the limit $r\to 1$ is not spoilt.

The subtraction and evaluation of the integrals can be performed analogously to the one-loop case
except that the integrations over $r$,$s$, and $t$ of the auxiliary integral must be performed
analytically.

\section{Applications}\label{sec:Applications}
To test our approach for the one-loop case, one can simply calculate single integrals and 
compare to \msbar-renormalized results, taking into account a possible finite part of the subtracted 
vacuum integral.\footnote{In \refe{Zirke:phd} this is shown to work for the massive bubble, 
where above threshold the auxiliary integral is analytically continued to the physical region, 
while the difference is integrated numerically using an appropriate contour deformation.}
At the two-loop level, however, this one-to-one correspondence between \fdr and \dr integrals 
is lost~\cite{Donati:2013voa}.
Thus we are forced to calculate physical observables in order to be able to compare to \dr results.

For this purpose we choose \nlo~\qcd corrections 
to the decay rate of scalar and pseudoscalar Higgs bosons into two photons in the heavy-top limit 
and to the $\rho$~parameter. 
Both quantities can be calculated with external momenta set to zero, 
\ie only vacuum integrals need to be evaluated.
Amongst other simplifications, this has the advantage that no thresholds or pseudo-thresholds 
will be present so we do not need to perform any analytic continuation or contour deformation
in this first test of our method.

Before presenting results for these observables in \scts{sec:Hgamgam} and \ref{sec:deltarho},
we briefly describe the setup of the calculation.\footnote{More details can be found in \refe{Zirke:phd}.}

\subsection{Setup}
\label{sec:setup}
In order to generate the amplitudes related to the observables we wish to compute,
we make use of the following tools:
\begin{itemize}
\item \qgraf~\cite{Nogueira:1991ex} for the generation of the diagrams,
\item \qtoe/\expp~\cite{Harlander:1997zb,Seidensticker:1999bb}
for topology matching, performing the asymptotic expansion~\cite{Smirnov:2002pj,Smirnov:1994tg} in small external momenta and
inserting the Feynman rules, and, where necessary,
\item \matad~\cite{Steinhauser:2000ry} to evaluate the vacuum integrals in \dr for comparison.
\end{itemize}

The next step is to interpret the integrals in the amplitude within the \fdr approach and 
derive the splitting into divergent and finite contributions,
which is done in an automated way in \form~\cite{Vermaseren:2000nd,Tentyukov:2007mu,Kuipers:2012rf} 
by systematic power counting and repeated application of \eqn{eq:FDRpartfrac}.
Our \form routines also introduce Feynman parameters and express the result in terms of
integrals of the type shown in \neqn{eq:int1} and~\eqref{eq:FDR2lfinal2}.
Factors involving the loop momentum in the numerator are taken into account by calculating
derivatives of intermediate Schwinger parameters~\cite{Smirnov:2006ry}, where the derivatives are performed
algebraically, \ie using relations of the type
 $\left[ \frac{\partial}{\partial x}, f(x) \right] = f'(x)$.

Then we switch to \mathem, which is used to evaluate the required auxiliary integrals case by case
for a given set of exponents, and write out the amplitude in terms of functions of the 
remaining variables as \cpp code.
To perform the numerical integrations we make use of the \cuba library~\cite{Hahn:2004fe}, 
where we choose the deterministic Cuhre algorithm~\cite{Berntsen:1991}, which turns out to perform best for the
type of functions that occur in our approach, at least in absence of thresholds.
To optimize the computation time we adjust the required precision for the individual integrals 
dynamically depending on their contribution to the final result.

\subsection[Higgs Decay into Two Photons at O(alpha alphaS)]{Higgs Decay into Two Photons at $\order{\alpha\alphas}$}
\label{sec:Hgamgam}
The decay of a Higgs boson into two photons in the limit of an infinite top-quark mass
already served as a test example for \fdr at the two-loop level in \refe{Donati:2013voa}, 
where agreement with the \dr result~\cite{Djouadi:1990aj,Dawson:1992cy} was found.
We will reproduce this result with our method and supplement it with the case of a pseudoscalar 
Higgs boson, which is of particular interest because it is linked to the 
axial anomaly~\cite{Adler:1969er}, as explained in \refe{Djouadi:1993ji}.

\begin{figure}[ht]
  \begin{center}
        \includegraphics[width=0.3\textwidth]{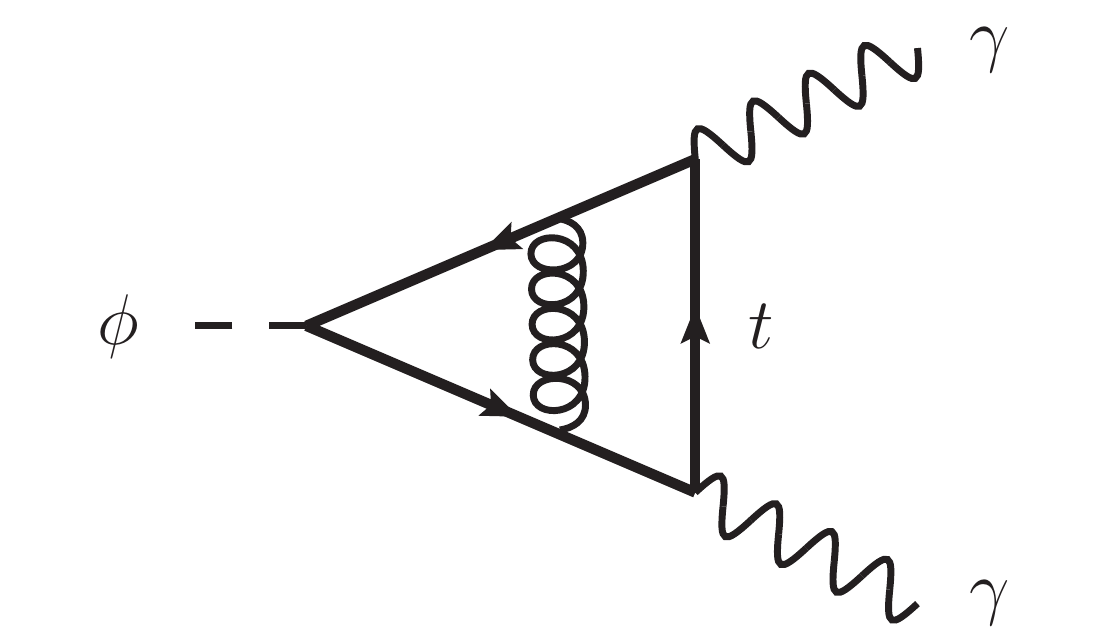}\quad
        \includegraphics[width=0.3\textwidth]{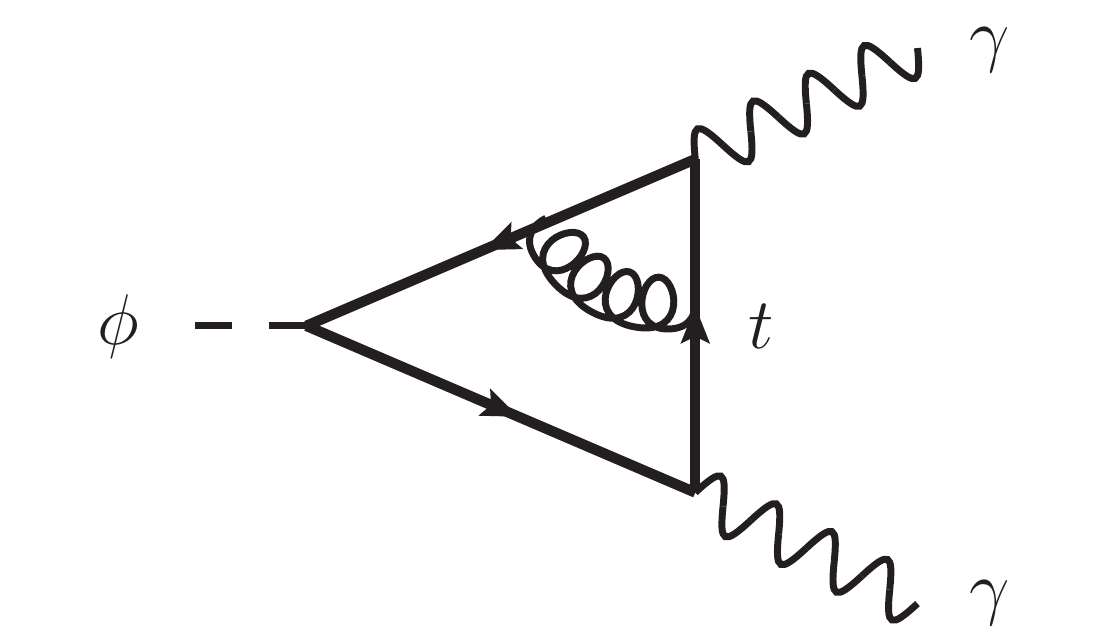}\quad
        \includegraphics[width=0.3\textwidth]{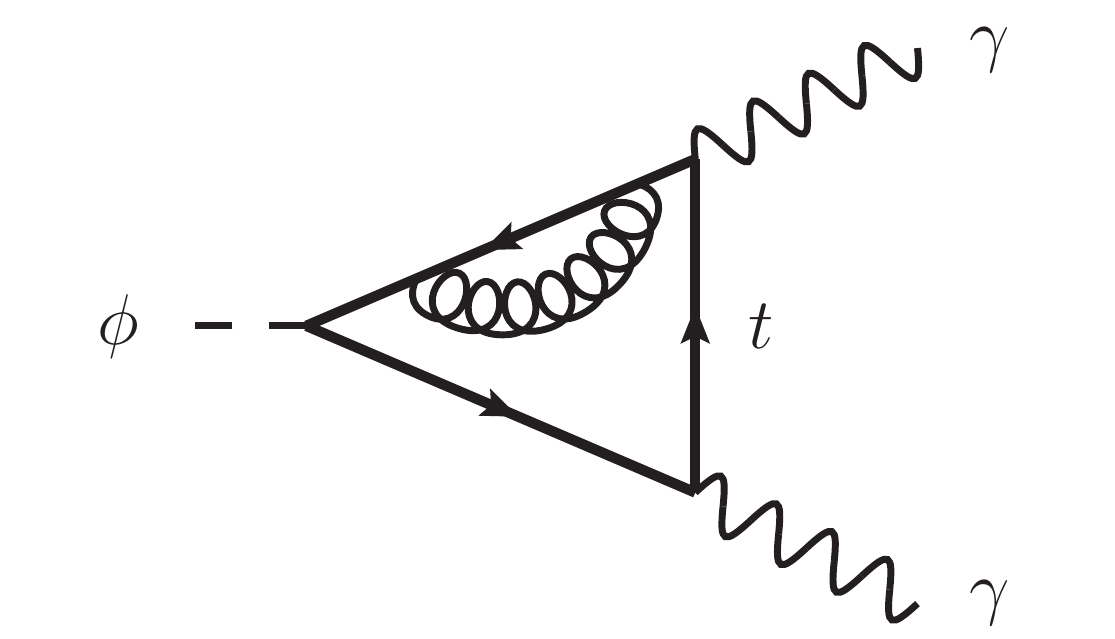}
  \end{center}
    \caption{Two-loop \qcd corrections to $\phi\rightarrow\gamgam$, $\phi\in\{h,A\}$.}
    \label{fig:phigamgamNLO}
\end{figure}
In terms of the momenta $q_{1,2}$ and the polarization vectors $\epsilon_{1,2}$ of the photons
the amplitude for the decay of a scalar (pseudoscalar) Higgs boson $h$ ($A$) can be written as
\begin{subequations}
\begin{align}
  \me_{h\rightarrow \gamgam} &= \left(\me_{h,t} + \me_{h,b} + \me_{h,W}\right) 
  \left( q_1\cdot q_2 \, \veps_1 \cdot \veps_2 - q_1\cdot\veps_2 \, q_2\cdot \veps_1\right)
  \; \text{or}\\
  \me_{A\rightarrow \gamgam}&= \left(\me_{A,t} + \me_{A,b}\right)
  \epsilon_{\mu\nu\rho\sigma}q_1^\mu q_2^\nu \veps_1^\rho \veps_2^\sigma,
\end{align}
\end{subequations}
respectively, and in general receives contributions from heavy quarks and, in the scalar case, $W$ bosons. 
Here we consider only \nlo~\qcd corrections to the top-loop contribution,
\begin{align}
\me_{\phi,t} = \me_{\phi,t}^{(0)} + \api \me_{\phi,t}^{(1)} + \order{\alphas^2}, \quad \phi\in\{h,A\},
\end{align}
for which typical diagrams are shown in \fig{fig:phigamgamNLO}.

Writing the top-quark couplings to the Higgs bosons generically as
\begin{subequations}
\begin{align}
\langle h t_i \bar{t}_j \rangle: &\quad i g_{ht\bar{t}} \frac{\mtbare}{v}\delta_{ij},\\
\langle A t_i \bar{t}_j \rangle: &\quad i g_{At\bar{t}} \frac{\mtbare}{v} \gamma_5 \delta_{ij},
\end{align}
\end{subequations}
the renormalized \dr amplitudes read\footnote{In the limit $\mt\to\infty$, the result is independent of the 
renormalization scheme chosen for $\mt$. For the case of the pseudoscalar Higgs, 
we treat $\gamma_5$ according to the scheme of Ref.~\cite{Larin:1993tq}, which requires
a finite renormalization to remove spurious axial anomalies.}
\begin{subequations}
  \label{eq:phigamgamDR}
\begin{align}
  \me_{h,t}(\mt\to\infty) &= -2\frac{\alpha}{\pi} Q_t^2 \frac{g_{ht\bar{t}}}{v} 
  \left( 1 - \api + \order{\alphas^2} \right), \\
  \me_{A,t}(\mt\to\infty) &= -3i\frac{\alpha}{\pi} Q_t^2 \frac{g_{At\bar{t}}}{v}
   \left( 1 + \order{\alphas^2} \right).
\end{align}
\end{subequations}

With the setup described in the previous section we obtain in \fdr 
(without any renormalization necessary)
\begin{subequations}
\begin{align}
  \frac{ \me_{h,t}^{(1)}(\mt\to\infty) }{\me_{h,t}^{(0)}(\mt\to\infty) } &= 
(1.9 \pm 3.3)\cdot 10^{-6} \frac{\mt^2}{s}  - \left\{ 1 + (2.7\pm 0.7) \cdot 10^{-6} \right\}, \\
  \frac{ \me_{A,t}^{(1)}(\mt\to\infty) }{\me_{A,t}^{(0)}(\mt\to\infty) } &= 
(0.1\pm 0.8) \cdot 10^{-6}, \label{eq:ampAgamgamfdr}
\end{align}
\end{subequations}
\ie we find numerical agreement to the level of $10^{-6}$. 
Note that in contrast to \dr the usage of $\gamma_5$ is unproblematic in \fdr. 
In the calculation of \eqn{eq:ampAgamgamfdr} we simply evaluated the fermion trace in four dimensions.

\subsection[Corrections to the rho Parameter at O(GF alphaS)]{Corrections to the $\rho$ Parameter at $\order{\GF\alphas}$}
\label{sec:deltarho}

\begin{figure}[ht]
\begin{center}
  \shortstack{
    \includegraphics[width=0.3\textwidth]{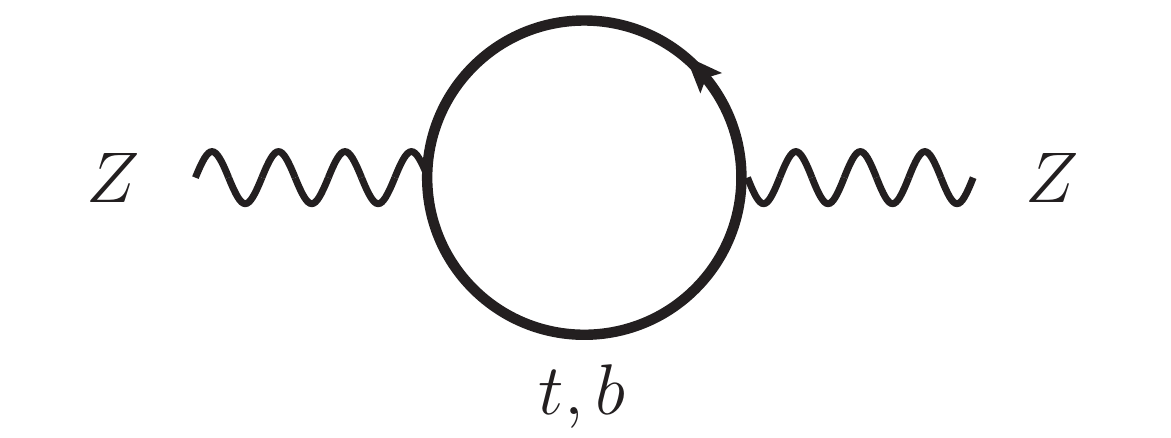}\quad
    \includegraphics[width=0.3\textwidth]{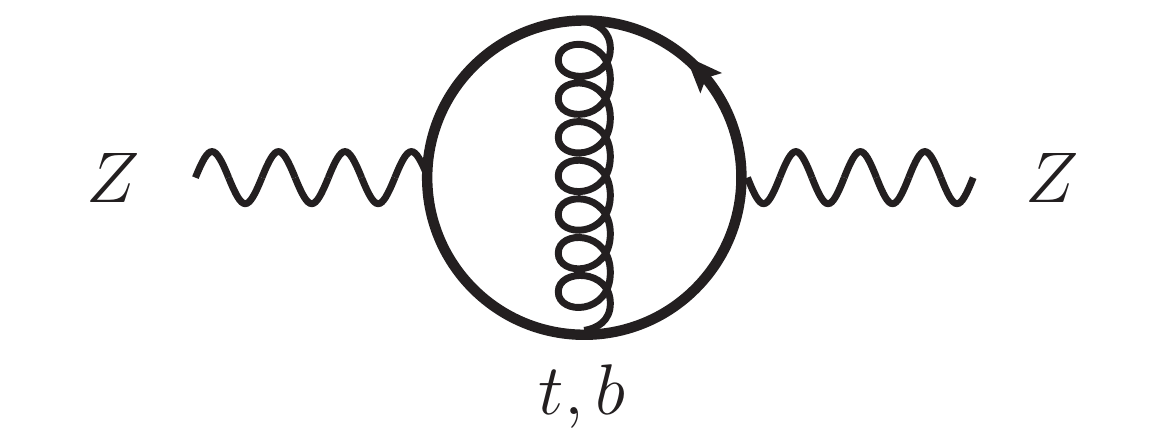}\quad
    \includegraphics[width=0.3\textwidth]{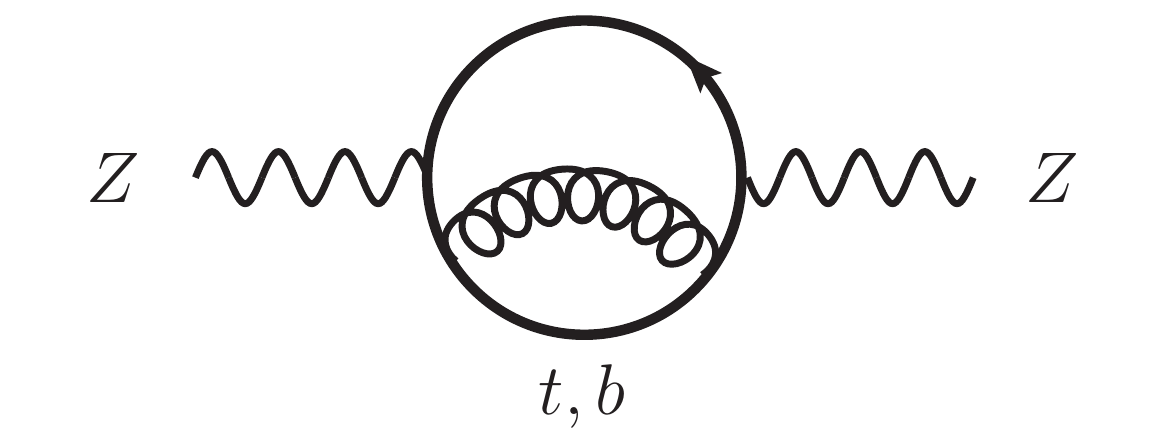}\\
    \includegraphics[width=0.3\textwidth]{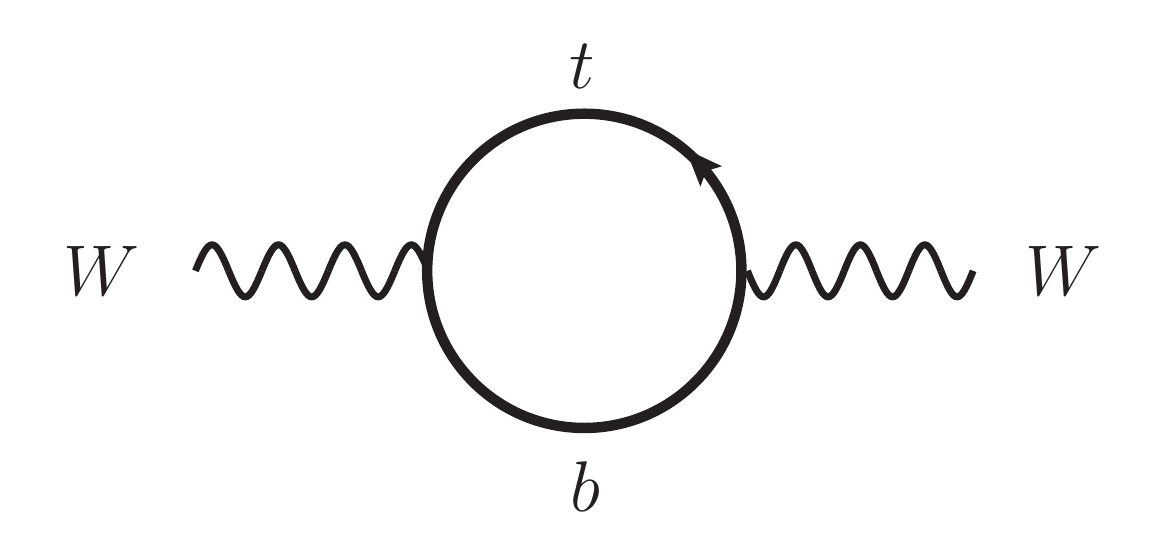}\quad
    \includegraphics[width=0.3\textwidth]{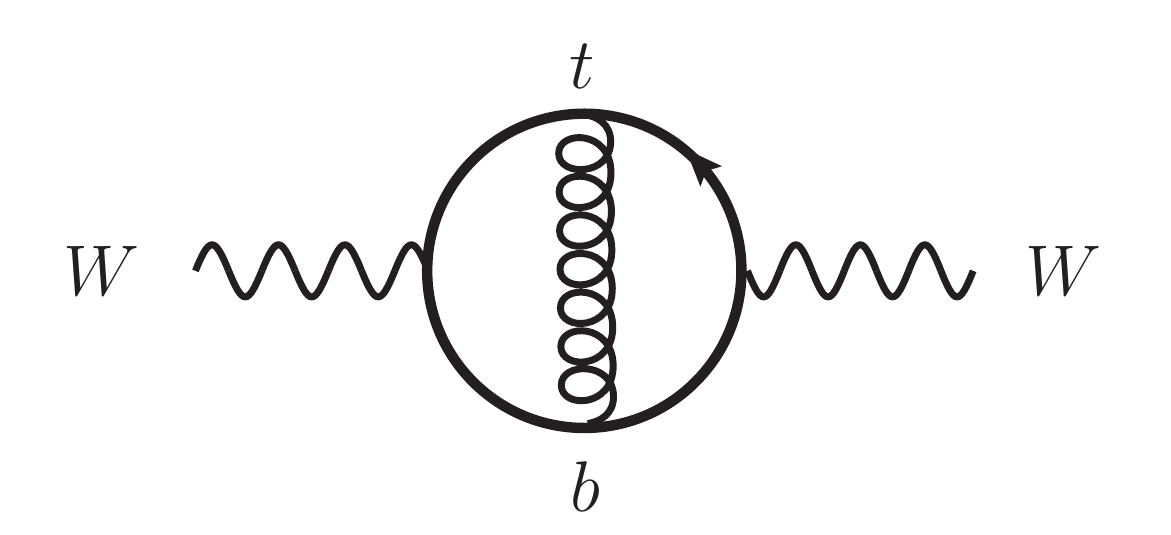}\quad
    \includegraphics[width=0.3\textwidth]{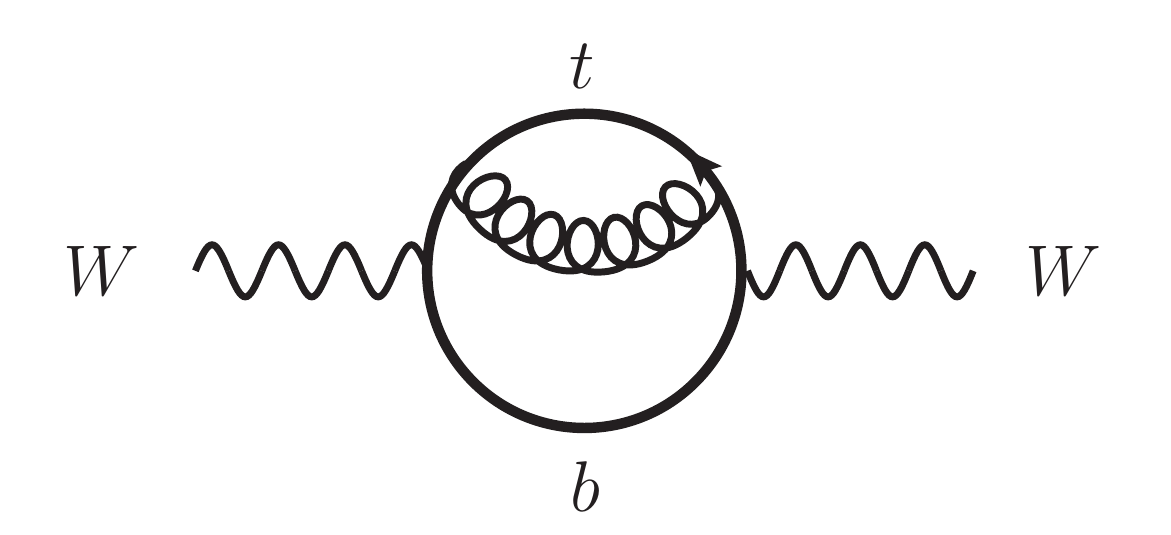}
  }
  \caption{Heavy-quark corrections to the $W$ and $Z$ propagator contributing at $\order{\GF}$ and 
    $\order{\GF\alphas}$.}
  \label{fig:WZpropdias}
\end{center}
\end{figure}
As another physical application we calculate the $\rho$~parameter~\cite{Ross:1975fq} to $\order{\GF\alphas}$ 
in \fdr. Its deviation from unity $\delta\rho$, defined by
\begin{equation}
  \rho = \frac{\mw^2}{\cos\thw^2\mz^2}= 1 + \delta\rho,
\end{equation}
can be expressed in terms of the transverse parts of the weak gauge boson polarization 
functions $\Pi_{VV}$, $V\in\{W,Z\}$, at zero momentum:
\begin{align}
  \delta\rho = \frac{\Pi_{ZZ}(0)}{\mz^2} - \frac{\Pi_{WW}(0)}{\mw^2}
\end{align}
Typical diagrams contributing up to $\order{\GF\alphas}$ are shown in \fig{fig:WZpropdias}.

The \dr result~\cite{Chetyrkin:1995ix} depends on the renormalization scheme for $\mt$ and reads
in the $\msbar$ and in the on-shell scheme
\begin{subequations}
  \label{eq:deltarhodr}
\begin{align}
  \delta\rho^{\msbar}&= \frac{3\GF\left(\mtmt\right)^2}{8\sqrt{2}\pi^2} \left[ 1 
    +\api \left( 2 - \frac 43 \zeta_2 + 2 L \right) + \order{\alphas^2}\right] \; \text{and}
  \label{eq:deltarhomsbar}\\
  \delta\rho^{\text{pole}}&= \frac{3\GF\left(\mtpole\right)^2}{8\sqrt{2}\pi^2} \left[ 1 
    +\api  \left(-\frac 23 - \frac 43\zeta_2 \right) + \order{\alphas^2} \right],
  \label{eq:deltarhopole}
\end{align}
\end{subequations}
respectively, where $L=\ln\left(\frac{\mu^2}{\mt^2}\right)$.

With our numerical \fdr setup we obtain at first
\begin{align}
\delta\rho^{\text{FDR}} &= \frac{3\GF\left(\mtFDR\right)^2}{8\sqrt{2}\pi^2}
\left[ 1 +\api \left( 2 L + 0.4734216(3) \right)\right] \nn
&= \frac{3\GF\left(\mtFDR\right)^2}{8\sqrt{2}\pi^2}
\left[ 1 +\api \left( 2 L + \frac 8 3 - \frac 43 \zeta_2 + (4\pm 3) \cdot 10^{-7}\right)\right],
\label{eq:deltarhoFDR}
\end{align}
where the top-quark mass is interpreted to be in the \fdr scheme, which is related to 
other schemes via a finite renormalization~\cite{Donati:2013voa,Page:2015zca}:
\begin{subequations}
\begin{align}
\frac{\mtFDR}{\mtmt} &= \frac{Z_m^\text{FDR}}{Z_m^\text{\msbar}} =
1 - \frac \alphas {3\pi} +\order{\alphas^2}, \\
\frac{\mtFDR}{\mtpole} &= 1 + \api \left(-L-\frac 5 3\right) +\order{\alphas^2}
\end{align}
\end{subequations}
Inserting either of these relations into \eqn{eq:deltarhoFDR} we find agreement
with \eqn{eq:deltarhodr} up to the level of $10^{-7}$.

\section{Conclusions}\label{sec:conclusions}
In this paper we have proposed a method to evaluate \ir-finite \fdr integrals numerically up to two-loop order
based on the subtraction of auxiliary integrals, which are constructed by linearizing parts of the denominator
in the Feynman parameter representation.
The method has been applied successfully to two-loop problems with vanishing external momenta,
adding the decay of a pseudoscalar Higgs boson into two photons in the heavy-top limit 
as well as \qcd corrections to the $\rho$~parameter to the list of observables recalculated in \fdr.
As expected, the treatment of $\gamma_5$ was found to be unproblematic in \fdr, 
at least for these examples.

The application to problems with finite external momenta, which is left for future investigations,
poses several challenges. Unfortunately, the relations~\eqref{eq:FDRpartfrac} lead to a large number of
tensor integrals in the presence of external momenta. Thus it is essential to reduce the number of integrals
as much as possible, which could be achieved by the application of integration-by-parts methods~\cite{Tkachov:1981wb,Chetyrkin:1981qh} 
as proposed in \refe{Pittau:2014tva}.
Furthermore, the introduction of additional massless propagators may increase the number of pseudo-thresholds 
with presumably disadvantageous effects on the stability of the numerical integration.

\paragraph{Acknowledgments.}
I would like to thank Robert Harlander for numerous fruitful discussions 
throughout this project.
The Feynman diagrams presented in this paper were created with the help of 
\jaxodraw~\cite{Vermaseren:1994je,Binosi:2003yf,Binosi:2008ig}.

\bibliographystyle{JHEP-mod}
\bibliography{References}

\providecommand{\href}[2]{#2}\begingroup\raggedright\begin{thebibliography}{10}

\bibitem{Pittau:2012zd}
R.~Pittau, \emph{A Four-Dimensional Approach to Quantum Field Theories},
  \href{http://dx.doi.org/10.1007/JHEP11(2012)151}{\emph{JHEP} {\bf 1211}
  (2012) 151}, [\href{http://arxiv.org/abs/1208.5457}{{\tt 1208.5457}}].

\bibitem{Donati:2013voa}
A.~M. Donati and R.~Pittau, \emph{FDR, an Easier Way to NNLO Calculations: A
  Two-Loop Case Study},
  \href{http://dx.doi.org/10.1140/epjc/s10052-014-2864-9}{\emph{Eur.~Phys.~J.}
  {\bf C74} (2014) 2864}, [\href{http://arxiv.org/abs/1311.3551}{{\tt
  1311.3551}}].

\bibitem{'tHooft:1972fi}
G.~'t~Hooft and M.~J.~G. Veltman, \emph{Regularization and Renormalization of
  Gauge Fields},
  \href{http://dx.doi.org/10.1016/0550-3213(72)90279-9}{\emph{Nucl.~Phys.} {\bf
  B44} (1972) 189--213}.

\bibitem{Pittau:2013qla}
R.~Pittau, \emph{QCD Corrections to $H \to gg$ in FDR},
  \href{http://dx.doi.org/10.1140/epjc/s10052-013-2686-1}{\emph{Eur.~Phys.~J.}
  {\bf C74} (2014) 2686}, [\href{http://arxiv.org/abs/1307.0705}{{\tt
  1307.0705}}].

\bibitem{Page:2015zca}
B.~Page and R.~Pittau, \emph{Two-loop off-shell QCD amplitudes in FDR},
  \href{http://dx.doi.org/10.1007/JHEP11(2015)183}{\emph{JHEP} {\bf 11} (2015)
  183}, [\href{http://arxiv.org/abs/1506.09093}{{\tt 1506.09093}}].

\bibitem{Dedes:2012hf}
A.~Dedes and K.~Suxho, \emph{Anatomy of the Higgs boson decay into two photons
  in the unitary gauge},
  \href{http://dx.doi.org/10.1155/2013/631841}{\emph{Adv. High Energy Phys.}
  {\bf 2013} (2013) 631841}, [\href{http://arxiv.org/abs/1210.0141}{{\tt
  1210.0141}}].

\bibitem{Donati:2013iya}
A.~M. Donati and R.~Pittau, \emph{Gauge Invariance at Work in FDR: $H \to
  \gamma \gamma$}, \href{http://dx.doi.org/10.1007/JHEP04(2013)167}{\emph{JHEP}
  {\bf 1304} (2013) 167}, [\href{http://arxiv.org/abs/1302.5668}{{\tt
  1302.5668}}].

\bibitem{Donati:2013bca}
A.~M. Donati and R.~Pittau, \emph{The $\gamma \gamma$ decay of the Higgs boson
  in FDR}, \href{http://dx.doi.org/10.1051/epjconf/20136012014}{\emph{EPJ Web
  Conf.} {\bf 60} (2013) 12014}, [\href{http://arxiv.org/abs/1306.6785}{{\tt
  1306.6785}}].

\bibitem{Nagy:2003qn}
Z.~Nagy and D.~E. Soper, \emph{General subtraction method for numerical
  calculation of one loop QCD matrix elements},
  \href{http://dx.doi.org/10.1088/1126-6708/2003/09/055}{\emph{JHEP} {\bf 09}
  (2003) 055}, [\href{http://arxiv.org/abs/hep-ph/0308127}{{\tt
  hep-ph/0308127}}].

\bibitem{Binoth:2000ps}
T.~Binoth and G.~Heinrich, \emph{An automatized algorithm to compute infrared
  divergent multiloop integrals},
  \href{http://dx.doi.org/10.1016/S0550-3213(00)00429-6}{\emph{Nucl. Phys.}
  {\bf B585} (2000) 741--759}, [\href{http://arxiv.org/abs/hep-ph/0004013}{{\tt
  hep-ph/0004013}}].

\bibitem{Heinrich:2008si}
G.~Heinrich, \emph{Sector Decomposition},
  \href{http://dx.doi.org/10.1142/S0217751X08040263}{\emph{Int. J. Mod. Phys.}
  {\bf A23} (2008) 1457--1486}, [\href{http://arxiv.org/abs/0803.4177}{{\tt
  0803.4177}}].

\bibitem{Zirke:phd}
T.~J.~E. Zirke, \emph{Perturbative Calculations and Their Application to Higgs
  Physics}.
\newblock PhD thesis, University of Wuppertal, 2014.

\bibitem{Smirnov:2006ry}
V.~A. Smirnov, \emph{Feynman Integral Calculus}.
\newblock Springer, 2006.

\bibitem{Soper:1999xk}
D.~E. Soper, \emph{Techniques for QCD calculations by numerical integration},
  \href{http://dx.doi.org/10.1103/PhysRevD.62.014009}{\emph{Phys. Rev.} {\bf
  D62} (2000) 014009}, [\href{http://arxiv.org/abs/hep-ph/9910292}{{\tt
  hep-ph/9910292}}].

\bibitem{Nagy:2006xy}
Z.~Nagy and D.~E. Soper, \emph{Numerical Integration of One-Loop Feynman
  Diagrams for $N$-Photon Amplitudes},
  \href{http://dx.doi.org/10.1103/PhysRevD.74.093006}{\emph{Phys.~Rev.} {\bf
  D74} (2006) 093006}, [\href{http://arxiv.org/abs/hep-ph/0610028}{{\tt
  hep-ph/0610028}}].

\bibitem{Nogueira:1991ex}
P.~Nogueira, \emph{Automatic Feynman Graph Generation},
  \href{http://dx.doi.org/10.1006/jcph.1993.1074}{\emph{J.~Comput.~Phys.} {\bf
  105} (1993) 279--289}.

\bibitem{Harlander:1997zb}
R.~Harlander, T.~Seidensticker and M.~Steinhauser, \emph{Complete Corrections
  of $\mathcal{O}(\alpha \alpha_s)$ to the Decay of the $Z$~Boson into Bottom
  Quarks},
  \href{http://dx.doi.org/10.1016/S0370-2693(98)00220-2}{\emph{Phys.~Lett.}
  {\bf B426} (1998) 125--132}, [\href{http://arxiv.org/abs/hep-ph/9712228}{{\tt
  hep-ph/9712228}}].

\bibitem{Seidensticker:1999bb}
T.~Seidensticker, \emph{Automatic Application of Successive Asymptotic
  Expansions of Feynman Diagrams},
  \href{http://arxiv.org/abs/hep-ph/9905298}{{\tt hep-ph/9905298}}.

\bibitem{Smirnov:2002pj}
V.~A. Smirnov, \emph{Applied Asymptotic Expansions in Momenta and Masses},
  {\emph{Springer Tracts Mod.~Phys.} {\bf 177} (2002) 1--262}.

\bibitem{Smirnov:1994tg}
V.~A. Smirnov, \emph{Asymptotic Expansions in Momenta and Masses and
  Calculation of Feynman Diagrams},
  \href{http://dx.doi.org/10.1142/S0217732395001617}{\emph{Mod.~Phys.~Lett.}
  {\bf A10} (1995) 1485--1500},
  [\href{http://arxiv.org/abs/hep-th/9412063}{{\tt hep-th/9412063}}].

\bibitem{Steinhauser:2000ry}
M.~Steinhauser, \emph{\texttt{MATAD}: A Program Package for the Computation of
  Massive Tadpoles},
  \href{http://dx.doi.org/10.1016/S0010-4655(00)00204-6}{\emph{Comput.~Phys.~Commun.}
  {\bf 134} (2001) 335--364}, [\href{http://arxiv.org/abs/hep-ph/0009029}{{\tt
  hep-ph/0009029}}].

\bibitem{Vermaseren:2000nd}
J.~Vermaseren, \emph{New Features of FORM},
  \href{http://arxiv.org/abs/math-ph/0010025}{{\tt math-ph/0010025}}.

\bibitem{Tentyukov:2007mu}
M.~Tentyukov and J.~A.~M. Vermaseren, \emph{The Multithreaded version of FORM},
  \href{http://dx.doi.org/10.1016/j.cpc.2010.04.009}{\emph{Comput. Phys.
  Commun.} {\bf 181} (2010) 1419--1427},
  [\href{http://arxiv.org/abs/hep-ph/0702279}{{\tt hep-ph/0702279}}].

\bibitem{Kuipers:2012rf}
J.~Kuipers, T.~Ueda, J.~A.~M. Vermaseren and J.~Vollinga, \emph{FORM version
  4.0}, \href{http://dx.doi.org/10.1016/j.cpc.2012.12.028}{\emph{Comput. Phys.
  Commun.} {\bf 184} (2013) 1453--1467},
  [\href{http://arxiv.org/abs/1203.6543}{{\tt 1203.6543}}].

\bibitem{Hahn:2004fe}
T.~Hahn, \emph{\texttt{CUBA}: A Library for Multidimensional Numerical
  Integration},
  \href{http://dx.doi.org/10.1016/j.cpc.2005.01.010}{\emph{Comput.~Phys.~Commun.}
  {\bf 168} (2005) 78--95}, [\href{http://arxiv.org/abs/hep-ph/0404043}{{\tt
  hep-ph/0404043}}].

\bibitem{Berntsen:1991}
J.~Berntsen, T.~O. Espelid and A.~Genz, \emph{An adaptive algorithm for the
  approximate calculation of multiple integrals},
  \href{http://dx.doi.org/10.1145/210232.210233}{\emph{ACM Trans. Math.
  Software} {\bf 17} (1991) 437--451}.

\bibitem{Djouadi:1990aj}
A.~Djouadi, M.~Spira, J.~J. van~der Bij and P.~M. Zerwas, \emph{QCD Corrections
  to $\gamma\gamma$ Decays of Higgs Particles in the Intermediate Mass Range},
  \href{http://dx.doi.org/10.1016/0370-2693(91)90879-U}{\emph{Phys.~Lett.} {\bf
  B257} (1991) 187--190}.

\bibitem{Dawson:1992cy}
S.~Dawson and R.~P. Kauffman, \emph{QCD corrections to $H\to \gamma \gamma$},
  \href{http://dx.doi.org/10.1103/PhysRevD.47.1264}{\emph{Phys.~Rev.} {\bf D47}
  (1993) 1264--1267}.

\bibitem{Adler:1969er}
S.~L. Adler and W.~A. Bardeen, \emph{Absence of higher order corrections in the
  anomalous axial vector divergence equation},
  \href{http://dx.doi.org/10.1103/PhysRev.182.1517}{\emph{Phys. Rev.} {\bf 182}
  (1969) 1517--1536}.

\bibitem{Djouadi:1993ji}
A.~Djouadi, M.~Spira and P.~M. Zerwas, \emph{Two Photon Decay Widths of Higgs
  Particles},
  \href{http://dx.doi.org/10.1016/0370-2693(93)90564-X}{\emph{Phys.~Lett.} {\bf
  B311} (1993) 255--260}, [\href{http://arxiv.org/abs/hep-ph/9305335}{{\tt
  hep-ph/9305335}}].

\bibitem{Larin:1993tq}
S.~A. Larin, \emph{The Renormalization of the Axial Anomaly in Dimensional
  Regularization},
  \href{http://dx.doi.org/10.1016/0370-2693(93)90053-K}{\emph{Phys.~Lett.} {\bf
  B303} (1993) 113--118}, [\href{http://arxiv.org/abs/hep-ph/9302240}{{\tt
  hep-ph/9302240}}].

\bibitem{Ross:1975fq}
D.~A. Ross and M.~J.~G. Veltman, \emph{Neutral Currents in Neutrino
  Experiments},
  \href{http://dx.doi.org/10.1016/0550-3213(75)90485-X}{\emph{Nucl.~Phys.} {\bf
  B95} (1975) 135}.

\bibitem{Chetyrkin:1995ix}
K.~G. Chetyrkin, J.~H. K\"uhn and M.~Steinhauser, \emph{Corrections of Order
  ${\cal O}(G_F M_t^2 \alpha_s^2)$ to the $\rho$ Parameter},
  \href{http://dx.doi.org/10.1016/0370-2693(95)00380-4}{\emph{Phys.~Lett.} {\bf
  B351} (1995) 331--338}, [\href{http://arxiv.org/abs/hep-ph/9502291}{{\tt
  hep-ph/9502291}}].

\bibitem{Tkachov:1981wb}
F.~V. Tkachov, \emph{A Theorem on Analytical Calculability of Four Loop
  Renormalization Group Functions},
  \href{http://dx.doi.org/10.1016/0370-2693(81)90288-4}{\emph{Phys. Lett.} {\bf
  B100} (1981) 65--68}.

\bibitem{Chetyrkin:1981qh}
K.~G. Chetyrkin and F.~V. Tkachov, \emph{Integration by Parts: The Algorithm to
  Calculate beta Functions in 4 Loops},
  \href{http://dx.doi.org/10.1016/0550-3213(81)90199-1}{\emph{Nucl. Phys.} {\bf
  B192} (1981) 159--204}.

\bibitem{Pittau:2014tva}
R.~Pittau, \emph{Integration-by-parts identities in FDR},
  \href{http://dx.doi.org/10.1002/prop.201500040}{\emph{Fortsch. Phys.} {\bf
  63} (2015) 601--608}, [\href{http://arxiv.org/abs/1408.5345}{{\tt
  1408.5345}}].

\bibitem{Vermaseren:1994je}
J.~A.~M. Vermaseren, \emph{Axodraw},
  \href{http://dx.doi.org/10.1016/0010-4655(94)90034-5}{\emph{Comput. Phys.
  Commun.} {\bf 83} (1994) 45--58}.

\bibitem{Binosi:2003yf}
D.~Binosi and L.~Theussl, \emph{JaxoDraw: A Graphical user interface for
  drawing Feynman diagrams},
  \href{http://dx.doi.org/10.1016/j.cpc.2004.05.001}{\emph{Comput. Phys.
  Commun.} {\bf 161} (2004) 76--86},
  [\href{http://arxiv.org/abs/hep-ph/0309015}{{\tt hep-ph/0309015}}].

\bibitem{Binosi:2008ig}
D.~Binosi, J.~Collins, C.~Kaufhold and L.~Theussl, \emph{JaxoDraw: A Graphical
  user interface for drawing Feynman diagrams. Version 2.0 release notes},
  \href{http://dx.doi.org/10.1016/j.cpc.2009.02.020}{\emph{Comput. Phys.
  Commun.} {\bf 180} (2009) 1709--1715},
  [\href{http://arxiv.org/abs/0811.4113}{{\tt 0811.4113}}].

\end{thebibliography}\endgroup

\end{document}